# Two-photon ionization of the *K*-shell of ions of the isonuclear sequence of a heavy atom

Alexey N. Hopersky, Alexey M. Nadolinsky, Sergey A. Novikov *and* Rustam V. Koneev

*Rostov State Transport University, 344038, Rostov-on-Don, Russia*
*E-mail*: qedhop@mail.ru, amnrnd@mail.ru, sanovikov@gmail.com, koneev@gmail.com

The shape and absolute values of the generalized cross-sections of the two-photon resonant single ionization of the *K*-shell of ions of the isonuclear sequence of a heavy nickel atom ($^{28}$Ni $\to$ Ni$^{26+}$ $\to$ Ni$^{24+}$ $\to$ Ni$^{18+}$) have been theoretically predicted. The complete wave functions of ionization states have been obtained in the single-configuration Hartree-Fock approximation. The effects of the occurrence of giant resonances in the subthreshold region of the generalized ionization cross-section and destructive quantum interference of the amplitudes of the probability of radiation transitions have been established. The leading role of the *d*-symmetry of the final state of ionization in determining the total generalized cross-section in the energy region of absorbed photons in the hard X-ray range has also been established.

## 1. Introduction

Two-photon resonance single ionization of the deep shell of an atom and ions of its isonuclear and isoelectron sequences by radiation from an X-ray free-electron laser (XFEL) is the subject of intensive experimental and theoretical research (see reviews [1, 2]). Within the framework of the second-order non-relativistic quantum perturbation theory, the first theoretical study of the generalized cross-sections of this process in the energy region of the ionization threshold of the *K*-shell (the region of birth of the resonance structure of the cross-section) of ions of the isonuclear sequence on the example of a light neon atom ($^{10}$Ne $\to$ Ne$^{8+}$ $\to$ Ne$^{6+}$) carried out in the authors' work [3]. In a recent article by the authors [4] on the example of a Ne-like ion of a heavy iron atom (Fe$^{16+}$) theory of work [3] modified to take into account, first of all, the full set of virtual (intermediate) states of photoexcitation, the non-trivial angular structure of the amplitudes of the probability of transitions to the final states of ionization *d*–symmetry and the effects of destructive quantum interference of amplitudes of the probability transitions. In this article, the theory of work [4] is generalized to ions of the isonuclear sequence of a heavy atom. As objects of research taken He-like (Ni$^{26+}$; ion ground state configuration and term $[0] = 1s^2$ [$^1S_0$]), Be-like (Ni$^{24+}$; $[0] = 1s^2 2s^2$ [$^1S_0$]) and Ne-like (Ni$^{18+}$; $[0] = 1s^2 2s^2 2p^6$ [$^1S_0$]) ions of the nickel atom ($^{28}$Ni; $[0] = 1s^2 2s^2 2p^6 3s^2 3p^6 3d^8 4s^2$ [$^3F_4$]). The choice of the isonuclear sequence is due to the spherical symmetry of the ground state of the ions, their availability in the gas phase [5] during the experiment on the absorption of two XFEL- photons by an ion trapped in the "trap" [6, 7] and the demand for their spectral characteristics, in particular, when studying hot laboratory [8] and astrophysical [9] plasma.

## 2. Theory

A detailed construction of the theory of two-photon resonant single ionization of the *K*-shell of an atomic ion with the $^1S_0$-term of the ground state (including the physical interpretation of ionization probability amplitudes in the representation of non-relativistic Feynman diagrams and taking into account the completeness of the set of virtual states of photoexcitation) is given in [4]. Without repeating these constructions, we will limit ourselves to the representation of two-photon ionization channels and the corresponding analytical structures of generalized ionization cross-sections for the ions under study.

Let us consider the two-photon ionization channel of the $K$-shell of ions $Ni^{26+}$ and $Ni^{24+}$:
$$2\omega + [0] \rightarrow 1s(n,x)p + \omega \rightarrow 1s\varepsilon l, \quad l = s, d. \tag{1}$$
For ion $Ni^{18+}$, along with the channel (1), channel defined:
$$2\omega + [0] \rightarrow 2p^5 xl + \omega \rightarrow 1s\varepsilon l. \tag{2}$$
Channel (2) realizes the effect of "reverse" X-ray emission ($1s \rightarrow 2p$ photoexcitation in the ion core)
$$1s^2 2p^5 + \omega \rightarrow 1s 2p^6, \tag{3}$$
where the electron of the continuous spectrum plays the role of an "observer". In (1) – (3) the atomic system of units is accepted ($e = \hbar = m_e = 1$), $\omega$ – energy of the absorbed photon, $x(\varepsilon)$ – continuous spectrum electron energy, $n$ – the principal quantum number of the excited state of the discrete spectrum and the filled shells of the ion configurations are not specified. The structure of the ionization channels presented corresponds to the following approximations. First. Strong energy separation of the $2s^2-$ and $2p^6-$ shells from $1s^2-$ shell of the ion $Ni^{18+}$ ($I_{1s}$ = 9059.70 eV (relativistic calculation of this work; see also [10]) $\gg I_{2s}$ ($I_{2p}$) = 1693.25 (1551.57) eV [11], $I_{nl}$ – ionization threshold energy of the $nl$ – ion shell) allows you to neglect the birth of finite $2s\varepsilon(s,d)-$ and $2p^5\varepsilon(p,f)-$ states of two-photon ionization at $\omega \gg I_{2s,2p}$. Second. Not counted in (2) intermediate $2p^5 nl$ –discrete spectrum states at $\omega \gg I_{2pnl}$ ($I_{2pnl}$ – energy $2p \rightarrow nl$ photoexcitation) are suppressed by the energy denominator $(\omega - I_{2pnl} + i\Gamma_{2p}/2)^{-1}$, where $\Gamma_{2p}$ is the width of the decay of $2p$-vacancies. Third. For ion $Ni^{24+}$ we have inequality $I_{1s}$ = 9991.42 eV (relativistic calculation of this work; see also [10]) $\gg I_{2s}$ = 2296.62 eV [12], which makes it possible to neglect the production of final $2s\varepsilon(s,d)$- states of two-photon ionization at $\omega \gg I_{2s}$. Fourth. The probability amplitude of two-photon ionization along the channel $2\omega + [0] \rightarrow 1s\varepsilon l$ determined by the contact interaction operator $\hat{C} \sim \sum_{n=1}^{N} (\hat{A}_n \cdot \hat{A}_n)$ ($N$ – the number of electrons in the ion, $\hat{A}_n$ – electromagnetic field operator in secondary quantization representation) and proportional to the matrix element $\langle 1s | j_l | \varepsilon l \rangle$, where $j_l$ – spherical Bessel function. In a dipole approximation for $\hat{A}_n$–operator (inequality is fulfilled $(r_{1s}/\lambda_\omega) \ll 1$, where $r_{1s}$ – medium radius $1s^2-$ shells and $\lambda_\omega$ – wavelength of the absorbed photon) have $j_0 \rightarrow 1$, $j_2 \rightarrow 0$ and $\langle 1s | j_l | \varepsilon l \rangle \rightarrow 0$ due to the orthogonality of the radial parts of wave functions $1s-$ and $\varepsilon l -$ states. For example, for ion $Ni^{26+}$ have: $r_{1s} = 0.029$ Å (calculation of this work), $\lambda_\omega = 1.024$ Å ($\omega = 10.3$ keV) and $r_{1s}/\lambda_\omega = 0.024 \ll 1$.

Let's consider the scheme of the proposed XFEL- experiment for linearly polarized photons: $\mathbf{k}$ (wave vector) $\in$ OZ, $\mathbf{e}_\omega$ (photon polarization vector) $\in$ OX, where OX, OZ – is the axes of the rectangular coordinate system. Then, according to [4], the analytical structure of the complete generalized cross-section of the two-photon ionization of the ion within the accepted approximations takes the form:

$$\sigma = (\eta/\omega^2) \sum_{l=s,d} \sum_{i=1,2} a_l L_{il}^2, \tag{4}$$

$$L_{1l} = \theta(\omega - \omega_{sp}) L_l + \sum_{n=3}^{\infty} (\omega - I_{1snp}) R_{ln}, \tag{5}$$

$$L_{2l} = \theta \gamma_{2p} L_l + D - \gamma_{1s} \sum_{n=3}^{\infty} R_{ln}, \qquad (6)$$

$$L_l = \frac{\omega_{sp}(2\omega - \omega_{sp})}{(\omega - \omega_{sp})^2 + \gamma_{2p}^2} \langle 1s_0 | \hat{r} | 2p_+ \rangle \langle 2p_0 | \hat{r} | \varepsilon_+ \rangle, \qquad (7)$$

$$R_{ln} = \frac{I_{1snp}(2\omega - I_{1snp})}{(\omega - I_{1snp})^2 + \gamma_{1s}^2} \cdot \langle 1s_0 \| \hat{r} \| np_+ \rangle \langle np_+ | \hat{r} | \varepsilon_+ \rangle, \qquad (8)$$

$$D = \sqrt{8\varepsilon} \cdot \langle 1s_0 \| \hat{r} \| \varepsilon p_+ \rangle, \qquad (9)$$

$$\langle 1s_0 \| \hat{r} \| np_+ \rangle = N_{1s} (\langle 1s_0 | \hat{r} | np_+ \rangle - \theta \cdot F_n), \qquad (10)$$

$$N_{1s} = \langle 1s_0 | 1s_+ \rangle \langle 2s_0 | 2s_+ \rangle^2 \langle 2p_0 | 2p_+ \rangle^6, \qquad (11)$$

$$F_n = \frac{\langle 1s_0 | \hat{r} | 2p_+ \rangle \langle 2p_0 | np_+ \rangle}{\langle 2p_0 | 2p_+ \rangle}, \qquad (12)$$

where $\eta = 0.278 \cdot 10^{-52}$ (cm$^4$·s), $a_s = 1$, $a_d = \frac{6}{5}\left(1 - \frac{1}{4\pi}\right)$, $\omega_{sp} = I_{1s} - I_{2p}$, $\varepsilon = 2\omega - I_{1s}$ ($\omega \geq I_{1s}/2$), $I_{1snp}$ – energy of the $1s \to np$ photoexcitation, $\gamma_{1s} = \Gamma_{1s}/2$, $\gamma_{2p} = \Gamma_{2p}/2$ and $\Gamma_{1s}$ ($\Gamma_{2p}$) – the width of the decay of the $1s$ ($2p$) – vacancies. In (7) – (12) indices "0" and "+" correspond to the radial parts of the wave functions of electrons obtained by solving the single-configuration equations of the self-consistent Hartree-Fock field for [0] and $1s_+\varepsilon l_+$ configurations of ion states. In (5), (6) and (10) the «control» $\theta$-parameter has been introduced: $\theta = 1$ for ion Ni$^{18+}$ and $\theta = 0$ for ions Ni$^{26+}$, Ni$^{24+}$. For $\theta = 0$ in (5) and (6) the summation index $n \in [2; \infty)$ and the equation structure (11) are modified: $N_{1s} = \langle 1s_0 | 1s_+ \rangle$ for ion Ni$^{26+}$ and $N_{1s} = \langle 1s_0 | 1s_+ \rangle \langle 2s_0 | 2s_+ \rangle^2$ for ion Ni$^{24+}$. For ions Ni$^{18+}$ and Ni$^{24+}$ the approximation of the independence of the parameter $\Gamma_{1s}$ from the main quantum number of the $1s \to np$ photoexcitation state is adopted. For ion Ni$^{26+}$ the following approximation of the radiative decay widths ($1snp$ ($^1P_1$) $\to 1s^2$($^1S_0$)) is adopted:

$$\Gamma_{1s} \to \Gamma_{1s,n} = \alpha n^{-\beta}, \qquad (13)$$

where the parameters $\alpha = 3.698$ and $\beta = 3.221$ determined according to the theoretical data of work [13] for $\Gamma_{1s,2}$ and $\Gamma_{1s,3}$. Taking into account (13), formulas (6) and (8) for ion Ni$^{26+}$ are modified:

$$\gamma_{1s} \sum_{n=2}^{\infty} \ldots \to \sum_{n=2}^{\infty} \gamma_{1s,n} \ldots, \quad \gamma_{1s}^2 \to \gamma_{1s,n}^2. \qquad (14)$$

## 3. Results and discussion

Table 1 shows the values of the parameters of the generalized cross-section (4) used in the calculations. In addition to Table 1, for ion Ni$^{18+}$ accepted values $\Gamma_{2p} = 0.036$ eV (extrapolation of data [13,16,17] for the widths of radiation decays $2p^5 n(s,d) \to 2p^6$, $n \in [3;\infty)]$) and $I_{2p}$ = 1551.57 eV [12]. For XFEL- photon energies accepted of the range $\hbar\omega \in (7.0; 11.5)$ keV (see [18] and references therein). The calculation results are shown in Figs. 1–3 and in Tables 2, 3.

The results in Figs. 1–3 and in Tables 2, 3 are presented in ordinary units (via Planck's constant) and demonstrate a pronounced subthreshold ($\hbar\omega \leq I_{1s}$) resonance structure of generalized cross-sections of two-photon ionization of the $K$-shell of ions Ni$^{18+}$, Ni$^{24+}$ and Ni$^{26+}$ through virtual states $1s \to np$ of photoexcitation (the values of the main quantum number $n_{\max}$

= 150 are taken into account). Giant resonance of a generalized cross-section at $\hbar\omega = 7.508$ keV for ion Ni$^{18+}$ (Fig. 1) is conditioned by the process $1s^2 2p^5 + \hbar\omega \to 1s 2p^6$ radiation absorption of the second photon incident on the ion. The magnitude of this resonance $\sigma \cong 4.058 \cdot 10^{-49}$ cm$^4 \cdot$s almost two orders of magnitude less than the value of the leading resonance $1s \to 3p$ photoexcitation, but significantly exceeds the values of photoexcitation $1s \to np$ resonances at $n \geq 4$ (Table 2).

Increase in generalized ionization cross-sections when switching from ion Ni$^{18+}$ to ions Ni$^{24+}$ and Ni$^{26+}$ (Tables 2, 3) is primarily due to a decrease in the width of decay $1s$–vacancy as a result of sequential "turning off" of the shielding effects $1s^2$–shell by the $2p^6$– и $2s^2$–shells. At the same time, the generalized ionization cross-section by virtual state $1s \to 2p$ photoexcitation for ion Ni$^{24+}$ exceeds that of ion Ni$^{26+}$ (Table 3). The reason is that in the denominator of the probability amplitude $R_{nl}$ of (8) for the corresponding $1s$- vacancy decay widths, the inequality is fulfilled 0.116 eV < 0.397 eV (Table 1). In fact, $2s^2$– shell of the ion Ni$^{24+}$ escapes $1s^2$–shell from virtual $2p$–state of photoexcitation, which leads to a decrease in the probability of radiation $1s 2p \to 1s^2$ decay compared to that for ion Ni$^{26+}$. As a result, the value $\Gamma_{1s}$ for the ion Ni$^{24+}$ is determined practically only by the probability of Auger-decay $1s 2s^2 \to 1s^2 xs$ in the presence of the excited $2p$-electron as an "observer". At the same time, the matrix elements of the radiation transition $1s \to 2p$ in (8) differ slightly: $\langle 1s_0 | \hat{r} | 2p_+ \rangle = 0{,}188$ (Ni$^{26+}$), 0.175 (Ni$^{24+}$). Of course, with an increase in the principal quantum number ($n \geq 3$), by virtue of formula (13), the values of the generalized cross-section resonances for the Ni$^{26+}$ ion exceed those for the Ni$^{24+}$ ion (Table 3). As a result, the near threshold structure of the generalized cross-section for the Ni$^{26+}$ ion (Fig. 3) becomes much more pronounced than that for the Ni$^{24+}$ ion (Fig. 2). It should be noted here that the shielding of the $1s^2$–shell by the $2s^2$- and $2p^6$– shells of the Ni$^{18+}$ ion is "compensated" by intense Auger-decays $1s$– vacancies involving $2s^2$– and $2p^6$ – shells and radiation decay $1s 2p^6 \to 1s^2 2p^5$ in the ion core.

The alternation of the $(\omega - \omega_{sp})$– and $(\omega - I_{1snp})$– factors in (5) leads to the effect of destructive (damping) quantum interference of the probability amplitudes of resonant radiation $1s^2 2p^5 \to 1s 2p^6$ and $1s \to np$ transitions (Figs. 1 – 3). As a result, wide "windows of transparency" occur between the maxima of generalized cross-section resonances as a sharp drop in the probability of two-photon ionization of the K-shell of Ni$^{18+}$, Ni$^{24+}$ and Ni$^{26+}$ ions.

## 4. Conclusion

A non-relativistic version of the quantum theory of the process of two-photon resonant single ionization of the K-shell of an isonuclear sequence of a heavy atom ion has been constructed. Ions with the $^1S_0$-term of the ground state are considered. As you might expect, the transition along the ion sequence is accompanied by pronounced quantitative and qualitative changes in the structure of the generalized ionization cross-sections. Taking into account correlational (going beyond the Hartree-Fock single-configuration approximation) and relativistic effects is the subject of future development of the theory. The results of successful experiments on the observation of two-photon ionization of atoms, molecules, and solids [1, 2, 20–25] suggest that the absolute values of the generalized cross-sections in Fig. 1 – 3 are quite measurable in the modern XFEL- experiment. The established subthreshold ($\hbar\omega \leq I_{1s}$) structure of generalized ionization cross-sections qualitatively reproduce the well-known «Fano profiles» [26,27]. In [28] it was shown that the two-photon absorption of circularly polarized radiation of the *optical* energy range by atoms of inert gases (He: $1s \to np$; Ar: $3p \to nd$; Xe: $3d \to nf$) allows the reproduction of subthreshold «Fano profiles» of photoelectron spectra with high spectral resolution. It is possible that the results of this work can be generalized to the *X-ray* range of the energies of photons absorbed by an atom (atomic ion).

**Table 1.** The decay widths of $1s$-vacancies ($\Gamma_{1s}$) and the energies of ionization thresholds $1s^2$–shells ($I_{1s}$) of $Ni^{26+}$, $Ni^{24+}$ and $Ni^{18+}$ ions.

| Ion | $\Gamma_{1s}$, eV | $I_{1s}$, eV |
|---|---|---|
| $Ni^{26+}$ | 0.397[a] | 10288.89[c] |
| $Ni^{24+}$ | 0.116[b] | 9991.42 |
| $Ni^{18+}$ | 1.246[b] | 9059.70 |

[a] Width of leading radiation decay $1s2p(^1P_1) \rightarrow 1s^2 (^1S_0)$ (theory of work [13]).
[b] Interpolation data of work [14].
[c] Relativistic calculation of work [15].

**Table 2.** Spectral characteristics of the leading $1s \rightarrow np$ resonances of the generalized cross-section of two-photon ionization of the $Ni^{18+}$ ion $K$-shell in the region of the energies of the absorbed photon $\hbar\omega \in (7.0; 9.5)$ keV. $[n] \equiv 10^n$.

| $np$ | $I_{1snp}$, keV | $\sigma$, $10^{-53}$ cm$^4\cdot$s |
|---|---|---|
| $3p$ | 8.432 | 1.119·[7] |
| $4p$ | 8.719 | 1.818·[4] |
| $5p$ | 8.845 | 5.263·[2] |

**Table 3.** See Table 2, but for ions $Ni^{24+}$, $Ni^{26+}$ at $\hbar\omega \in (7.0; 11.5)$ keV.

| Ion | $np$ | $I_{1snp}$, keV | $\sigma$, $10^{-47}$ cm$^4\cdot$s |
|---|---|---|---|
| $Ni^{24+}$ | $2p$ | 7.706 | 4.144·[6] |
| | $3p$ | 9.010 | 0.682·[4] |
| | $4p$ | 9.450 | 0.171·[3] |
| $Ni^{26+}$ | $2p$ | 7.784 | 5.363·[5] |
| | $3p$ | 9.177 / 9.184[a] | 1.298·[4] |
| | $4p$ | 9.664 / 9.667[a] | 2.135·[3] |

[a] Relativistic calculation of work [19].

**Table 4.** The relative contribution of the $s-$ and $d-$symmetries of the final ionization state $\zeta = \sigma_d/\sigma_s$ (see (4) for $l = s, d$) to the total generalized cross-section of the two-photon ionization of the $K$–shell of the $Ni^{18+}$, $Ni^{24+}$ and $Ni^{26+}$ ions.

| Ion | $\hbar\omega$, keV | $\zeta$ |
|---|---|---|
| $Ni^{18+}$ | 9.060 | 2.751 |
| $Ni^{24+}$ | 9.995 | 2.907 |
| $Ni^{26+}$ | 10.300 | 2.917 |

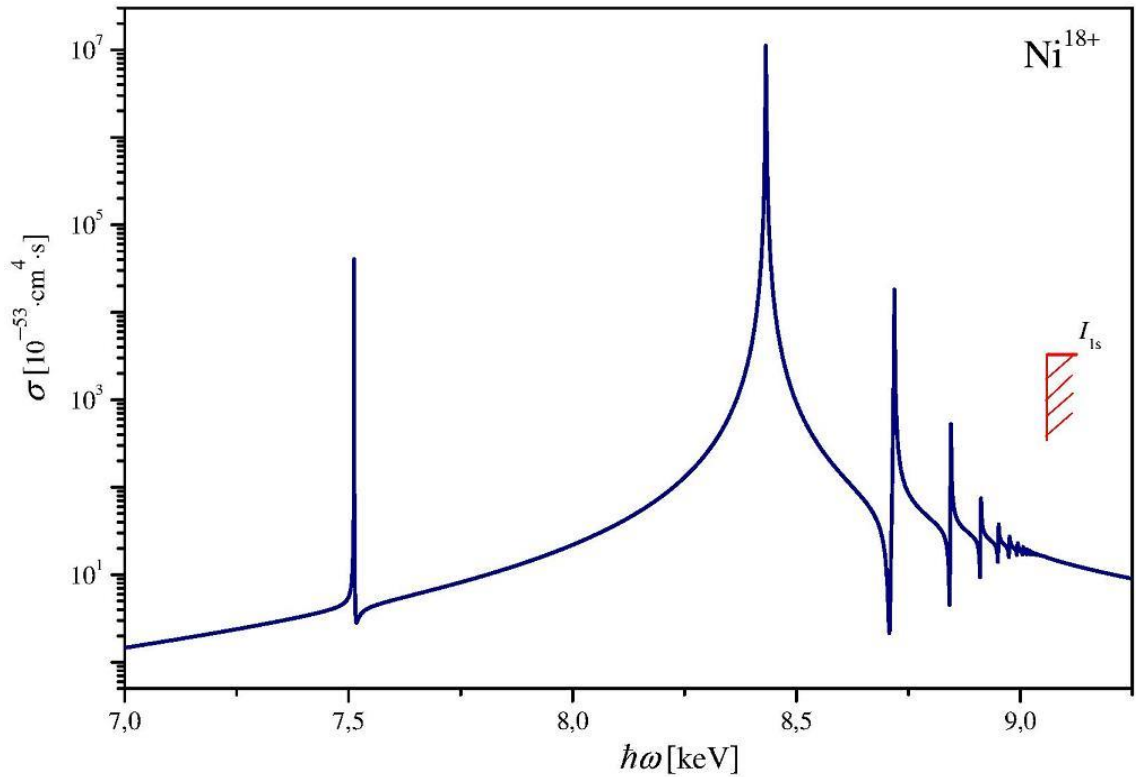

**Figure 1.** The total generalized cross-section of the two-photon resonant single ionization of the $K$-shell of the Ni$^{18+}$ ion. $\hbar\omega$ is the energy of the absorbed photon. $I_{1s}$ – energy of the ionization threshold of the $1s^2$ – shell.

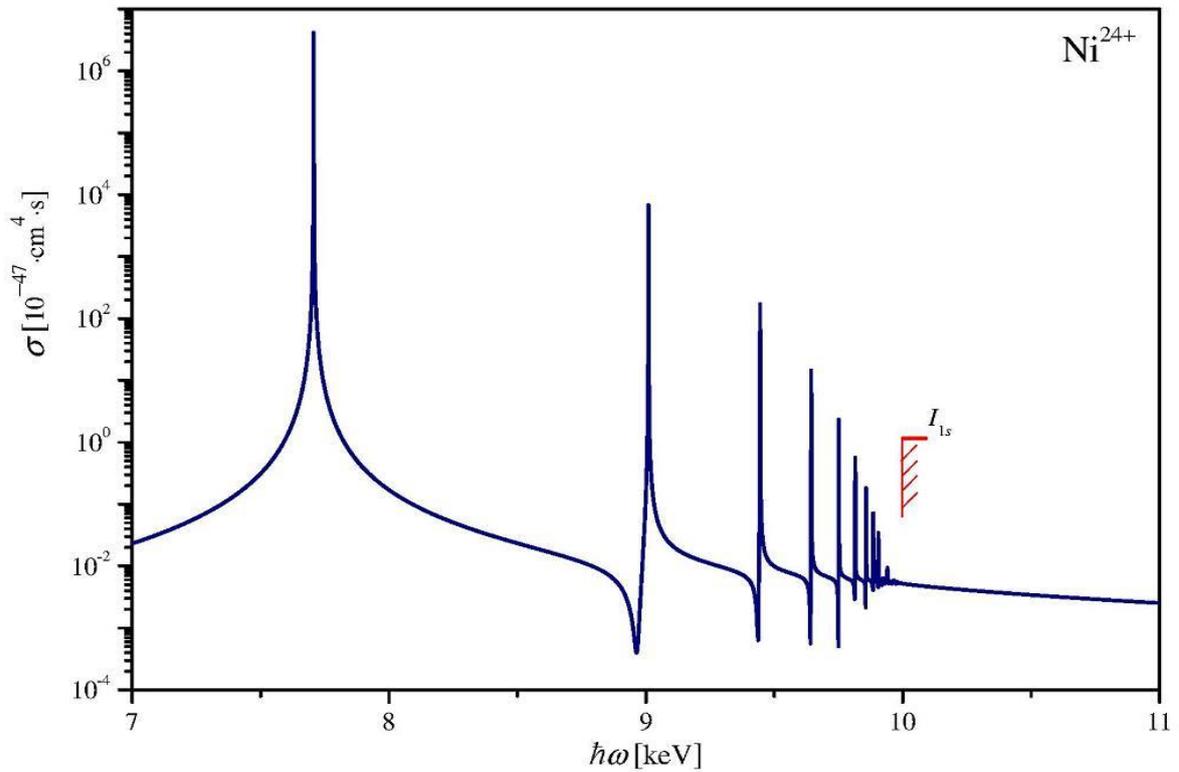

**Figure 2.** See Fig. 1, but for the Ni$^{24+}$ ion.

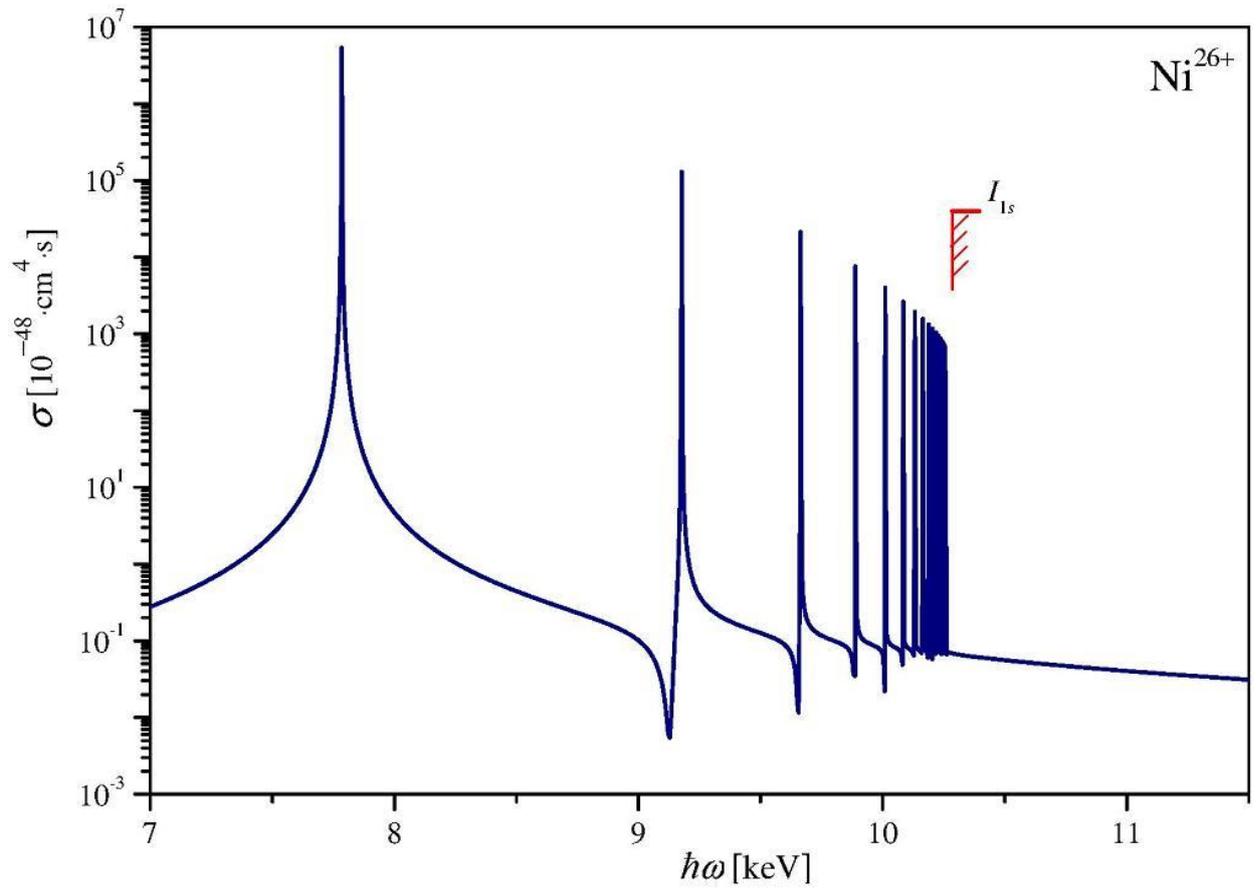

**Figure 3.** See Fig. 1, but for the Ni$^{26+}$ ion.